\documentstyle[12pt]{article}

\def\hybrid{\topmargin 0pt      \oddsidemargin 0pt
        \headheight 0pt \headsep 0pt
        \voffset=-0.5cm
        \textwidth 6.5in        
        \textheight 9in         
        \marginparwidth 0.0in
        \parskip 5pt plus 1pt   \jot = 1.5ex}
\catcode`\@=11
\def\marginnote#1{}

\newcount\hour
\newcount\minute
\newtoks\amorpm
\hour=\time\divide\hour by60
\minute=\time{\multiply\hour by60 \global\advance\minute by-\hour}
\edef\standardtime{{\ifnum\hour<12 \global\amorpm={am}%
        \else\global\amorpm={pm}\advance\hour by-12 \fi
        \ifnum\hour=0 \hour=12 \fi
        \number\hour:\ifnum\minute<10 0\fi\number\minute\the\amorpm}}
\edef\militarytime{\number\hour:\ifnum\minute<10 0\fi\number\minute}

\def\draftlabel#1{{\@bsphack\if@filesw {\let\thepage\relax
   \xdef\@gtempa{\write\@auxout{\string
      \newlabel{#1}{{\@currentlabel}{\thepage}}}}}\@gtempa
   \if@nobreak \ifvmode\nobreak\fi\fi\fi\@esphack}
        \gdef\@eqnlabel{#1}}
\def\@eqnlabel{}
\def\@vacuum{}
\def\draftmarginnote#1{\marginpar{\raggedright\scriptsize\tt#1}}

\def\draftlabel#1{{\@bsphack\if@filesw {\let\thepage\relax
   \xdef\@gtempa{\write\@auxout{\string
      \newlabel{#1}{{\@currentlabel}{\thepage}}}}}\@gtempa
   \if@nobreak \ifvmode\nobreak\fi\fi\fi\@esphack}
        \gdef\@eqnlabel{#1}}
\def\@eqnlabel{}
\def\@vacuum{}
\def\draftmarginnote#1{\marginpar{\raggedright\scriptsize\tt#1}}

\def\draft{\oddsidemargin -.5truein
        \def\@oddfoot{\sl preliminary draft \hfil
        \rm\thepage\hfil\sl\today\quad\militarytime}
        \let\@evenfoot\@oddfoot \overfullrule 3pt
        \let\label=\draftlabel
        \let\marginnote=\draftmarginnote
   \def\@eqnnum{(\theequation)\rlap{\kern\marginparsep\tt\@eqnlabel}%
\global\let\@eqnlabel\@vacuum}  }

\def\numberbysection{\@addtoreset{equation}{section}
        \def\theequation{\thesection.\arabic{equation}}}

\def\underline#1{\relax\ifmmode\@@underline#1\else
        $\@@underline{\hbox{#1}}$\relax\fi}

\def\titlepage{\@restonecolfalse\if@twocolumn\@restonecoltrue\onecolumn
     \else \newpage \fi \thispagestyle{empty}\c@page\z@
        \def\thefootnote{\fnsymbol{footnote}} }

\def\endtitlepage{\if@restonecol\twocolumn \else  \fi
        \def\thefootnote{\arabic{footnote}}
        \setcounter{footnote}{0}}  
\relax

\numberbysection
\hybrid

\def\beq{\begin{equation}}
\def\eeq{\end{equation}}
\def\p{\partial}
\def\G{\Gamma}
\def\g{\gamma}
\def\s{\sigma}
\def\z{\zeta}

\def\tL{\tilde L}
\def\C{{\cal C}}
\def\tC{\tilde {\cal C}}
\def\a{\alpha}

\def\f{\varphi}

\def\M{{\cal M}}

\def\res{{\rm res}}

\def \matrix #1 {\left(\begin{array}{cc} #1 \end{array}\right)}
\newtheorem{th}{Theorem}[section]

\newtheorem{cor}{Corollary}[section]
\newtheorem{lem}{Lemma}[section]

\begin{document}

\begin{titlepage}
\title{Elliptic analog of the Toda lattice}

\author{I.Krichever \thanks{Columbia University, 2990 Broadway,
New York, NY 10027, USA and
Landau Institute for Theoretical Physics,
Kosygina str. 2, 117940 Moscow, Russia; e-mail:
krichev@math.columbia.edu}}
\date{}

\maketitle

\begin{abstract} The action-angle variables for $N$-particle
Hamiltonian system with the Hamiltonian
$H=\sum_{n=0}^{N-1} \ln sh^{-2}\left(p_n/2\right)+\ln\left(\wp(x_n-x_{n-1})-
\wp(x_n+x_{n-1})\right), \ x_N=x_0,$ are constructed, and the system is solved
in terms of the Riemann $\theta$-functions. It is shown that
this system describes pole dynamics of the elliptic solutions
of 2D Toda lattice corresponding to spectral curves defined by the equation
$w^2-P_{N}^{el}(z)w+\Lambda^{2N}=0$, where $P_{N}^{el}(z)$ is an elliptic
function with pole of order $N$ at the point $z=0$.

\end{abstract}

\vfill

\end{titlepage}
\newpage

\section{Introduction}
The main goal of this paper is to construct the
action-angle variables for a finite dimensional Hamiltonian
system of equations
\beq
\ddot x_{n}=(\dot x_{n}^2-1) (V(x_n,x_{n+1})+V(x_n,x_{n-1}),\ \ x_{n+N}=x_n,
\label{sys}
\eeq
where
\beq\label{sys1}
V(u,v)=\z(u-v)+\z(u+v)-\z(2u)=-\ {1\over 2}\ {\wp'(u-v)-\wp'(u+v)\over
\wp(u-v)-\wp(u+v)}\ ,
\eeq
and to identify it as {\it an elliptic analog of $N$-periodic Toda lattice}.
Here $\wp(x)=\wp(x|2\omega,2\omega')$ and $\zeta(x)=\zeta(x|2\omega,2\omega')$
are classical Weierstrass functions.

Recently, finite-dimensional integrable soliton systems have
attracted very special interest due to their unexpected relations to
the theory of supersymmetric gauge models. The celebrated Seiberg-Witten
ansatz (\cite{sw1,sw2}) identifies moduli space of physically non-equivalent
vacua of the model with moduli space of a certain family of algebraic curves.
In \cite{af,klem} it was shown that the family of curves corresponding to
four-dimensional $N=2$ supersymmetric $SU(N_c)$ theory is defined by
the equation
\beq
w^2-wP_{N_c}(E)+\Lambda^{2N_c}=0,
\ \ P_{N_c}(E)=E^{N_c}+\sum_{i=0}^{N_c-1}u_iE^i.
\label{fam}
\eeq
In \cite{gor} it was noted that this family can be identified with the family
of spectral curves of $N_c$-periodic Toda lattice, and
the Seiberg-Witten ansatz  was linked with the Whitham perturbation theory
of finite-gap solutions of soliton equations proposed in
\cite{kr1,kr2}. Integrable systems related to various gauge models coupled
with matter hypermultiplets in various representations were considered
in \cite{mar}-\cite{mm2}, where more complete list of references
can be found.

In \cite{ggm1,mm2} $N_c$-periodic spin chain related to $XYZ$ model was
proposed as soliton counterpart of $N=1$ supersymmetric $SU(N_c)$ theory
in six dimensions compactified in two directions, and coupled with
$N_f=2N_c$ matter hypermultiplets.
Spectral curves of $N_c$-periodic homogeneous $XYZ$ spin chain have the form
\beq\label{xyz}
w^2-wP_{N_c}^{el}(z)+Q_{2N_c}^{el}(z)=0,
\eeq
where $P_{N_c}^{el}(z)$ and $Q_{2N_c}^{el}(z)$ are elliptic polynomials, i.e.
elliptic functions with poles of order $N_c$ and $2N_c$ at the point $z=0$.
Note, that (\ref{xyz}) is an elliptic deformation of the family of
curves found in \cite{ho} for four-dimensional $N=2$ sypersymmetric
$SU(N_c)$ model coupled with matter hypermultiplets.

A particular case of (\ref{xyz}), when $Q_{2N_c}^{el}(z)$ is a constant, $Q_{2N_c}^{el}(z)=\Lambda^{2N_c}$ can be seen as an elliptic deformation
of (\ref{fam}).
The corresponding family of curves depends on $N_c$ parameters which
can be chosen as $\Lambda$ and the coefficients $u_i$ of the representation
of $P_N^{el}(z)$ in the form:
\beq\label{el}
P_N^{el}(z)={(-1)^{N}\over (N-1)!}
\p_z^{N-2}\wp(z)+\sum_{i=1}^{N-2}u_i\p_z^{i-1}\wp(z)+u_0,
\eeq
An attempt to find a soliton system corresponding to the family
of spectral curves defined by the equation
\beq
w^2-wP_N^{el}(z)+\Lambda^{2N}=0,
\label{fam1}
\eeq
led us to (\ref{sys}). After the system was found it turned out
that, by itself, it is not new. Up to a change of variables $q_n=\wp(x_n)$,
it coincides with one of the systems listed in \cite{shab}, where the
classification of all Toda type chains which have Toda type symmetries
was obtained. The new results obtained in this work are: isomorphism of
(\ref{sys}) with a pole system corresponding to elliptic solutions of
2D Toda lattice, the construction of action-angle variables, and
explicit solution of the system in terms of the theta-functions.

In \cite{kp1,kp2} it was shown that a wide class of solutions of
the Seiberg-Witten ansatz can be described in terms of a special foliation
on the moduli space of curves with punctures. That allows one to consider
such systems as reductions of 2D soliton equations.
Following this approach, let us note, that (\ref{fam1})
defines an algebraic curve $\G$ as two-sheeted
cover of the elliptic curve $\G_0$ with periods $2\omega,2\omega'$.
Let $P_{\pm}$ be preimagies on $\G$ of $z=0$.
According to the construction of \cite{kr3}, any algebraic curve with two
punctures generates a family of algebro-geometric
solutions of 2D Toda lattice
\beq \label{toda}
(\p^2_{tt}-\p^2_{xx})\f_n=4\left(e^{\f_{n+1}-\f_n}-e^{\f_{n}-\f_{n-1}}\right),
\eeq
parameterized by points of the Jacobian $J(\G)$ of the curve.

In the next section we show that algebro-geometric solutions $\f_n(x,t)$
corresponding to $\G$ defined by (\ref{fam1}) are periodic in $n$ up to the
shift, $\f_n=\f_{n+N}+2N\ln \Lambda$, and have the form
\beq \label{f}
\f_n(x,t)=\a_n(t)+\ln {\s(x-x_{n+1}(t)+a)\s(x+x_{n+1}(t)+a)\over
\s(x-x_{n}(t)+a)\s(x+x_{n}(t)+a)}\ .
\eeq
Substitution of (\ref{f}) into (\ref{toda}) leads to equations (\ref{sys})
for $x_n(t)$.

It section 3 we construct a new Lax representation for (\ref{sys}) and
show that the spectral curve defined by the Lax operator has the form
(\ref{fam1}). We prove also, that if $x_n(t)$ is a solution of
(\ref{sys}), then there exist functions $\a_n(t)$ (unique up
to the transformation $\a_n(t)\to \a_n(t)+c_1t+c_2, \ c_i=const$), such that
the functions $\f_n(x,t)$ of the form (\ref{f}) satisfy (\ref{toda}).

The last section is devoted to the Hamiltonian theory of system (\ref{sys}).
Equations (\ref{sys}) are generated by the Hamiltonian
\beq
H=\sum_{n=0}^{N-1} \ln sh^{-2}\left(p_n/2\right)+\ln\left(\wp(x_n-x_{n-1})-
\wp(x_n+x_{n-1})\right),
\label{ham}
\eeq
and the canonical Poisson brackets $\{p_m,x_n\}=\delta_{nm}$.
We would like to emphasize that though this Hamiltonian structure
can be easily checked directly, it was found by the author using
the algebro-geometric approach to Hamiltonian theory of the Lax
equations proposed in \cite{kp1,kp2}, and developed in \cite{kr4}.
The main advantage of this approach is that it allows us to find simultaneously
the action-angle variables and a generating differential
which defines low-energy effective prepotential.

Note, that from the relation of system (\ref{sys}) to 2D Toda lattice
it is clear that degeneration of the elliptic curve $\G_0$ corresponds to
a degeneration of this system to the Toda lattice. It would be very
interesting to consider this degeneration explicitly on the level of
the Hamiltonian structure. We will consider this problem elsewhere.

\section{Elliptic solutions of 2D Toda lattice}

Algebro-geometric solutions of 2D Toda lattice were constructed in \cite{kr3}.
Let $\G$ be a smooth genus $g$ algebraic curve with fixed
local coordinates $z_{\pm}(Q)$ in neighborhoods of two punctures
$P_{\pm}\in \G,\ z_{\pm}(P_{\pm})=0$. Then, for any set of $g$ points
$\g_1,\ldots,\g_g$ in general position there exists a unique function
$\psi_n(x,t,Q)$ such that:

$1^0.$ $\psi_n(x,t,Q)$, as a function of the variable $Q\in \G$, is
meromorphic on $\G$ outside the punctures $P_{\pm}$ and has at most simple
poles at the points $\g_s$ (if all of them are distinct);

$2^0.$ in the neighborhoods of the punctures the function $\psi_n$ has the
form
\beq\label{1}
\psi_n=z_{\pm}^{{\mp}N} e^{(x\pm t)z^{-1}}
\left(\sum_{s=0}^{\infty}\xi^{\pm}_s(x,t)z_{\pm}^s\right),\ \ \xi_0^+=1.
\eeq
Uniqueness of $\psi_n$ implies that it satisfies the following system of
linear equations
\begin{eqnarray}
(\p_t+\p_x)\psi_n(x,t,Q)&=&2\psi_{n+1}(x,t,Q)+v_n(x,t)\psi_n(x,t,Q),
\label{2}\\
(\p_t-\p_x)\psi_n(x,t,Q)&=&2c_n(x,t)\psi_{n-1}(x,t), \label{3}
\end{eqnarray}
where the coefficients are defined by the leading coefficient $\xi_0^-$
of expansion (\ref{1}) with the help of the formulae:
\beq
v_n=(\p_t+\p_x)\f_n(x,t),\ \ c_n=e^{\f_n(x,t)-\f_{n-1}(x,t)},\ \
\f_n(x,t)=\ln \xi_0^-(x,t).
\eeq
Compatibility of (\ref{2}) and (\ref{3}) implies that
$\f_n(x,t)$ is a solution of 2D Toda lattice  (\ref{toda}).

The function $\psi_n(x,t,Q)$ is called the Baker-Akhiezer function and
can be explicitly expressed in terms of the Riemann theta-function
associated with a matrix of $b$-periods of holomorphic differentials on $\G$.
The corresponding formula for $\f_n$ is as follows.

Let us fix a basis of cycles $a_i,b_i,\ i=1,\ldots,g,$ on $\Gamma$ with the
canonical matrix of intersections: $a_i\circ a_j=b_i\circ b_j=0,
\ a_i\circ b_j=\delta_{ij}$. The basis of normalized
holomorphic differentials $d\Omega^h_j(Q),\ j=1,\ldots,g,$ is defined by
conditions
$
\oint_{a_i}d\Omega_j^h= \delta_{ij} .
$
The $b$-periods of these differentials define the Riemann matrix
$
B_{kj}=\oint_{b_j} d\Omega_k^h .
$
The basic vectors $e_k$ of $C^g$ and the vectors $B_k$, which are
columns of the matrix $B$, generate a lattice ${\cal B}$ in $C^g$.  The
$g$-dimensional complex torus
\beq J(\Gamma)=C^g/{\cal B}, \ \
{\cal B}=\sum n_k e_k+m_k B_k,\ \ n_k,m_k \in Z,  \label{39}
\eeq
is called the Jacobian variety of $\Gamma$. A vector with the coordinates
\beq\label{ab}
A_k(Q)=\int_{P_+}^Q d\Omega_k^h
\eeq
defines the Abel map
$A: \Gamma \longrightarrow J(\Gamma).$

The Riemann matrix has a positive-definite imaginary part.
The entire function of $g$ variables $z=(z_1,\ldots,z_g)$
$$
\theta (z) =\theta (z\vert B)=\sum_{m\in Z^g} e^{2\pi i (z,m)+ \pi i (Bm,m)},
$$
is called the Riemann theta-function. It has the following
monodromy properties
\beq
\theta (z+e_k)=\theta (z),\ \ \theta (z+B_k)=e^{-2\pi i z_k- \pi i B_{kk}}
\theta (z). \label{5}
\eeq
The function $\theta (A(Q)+Z)$ is a multi-valued function of $Q$. But
according
to (\ref{5}), the zeros of this function are well-defined. For $Z$ in a
general position the equation
$
\theta(A(Q)+Z)=0
$
has $g$ zeros $\gamma_1,\ldots,\gamma_g$.
The vector $Z$ and the divisor of these zeros are connected by the relation
$
Z=-\sum_{s}A(\gamma_s)+{\cal K},\label{44}
$
where ${\cal K}$ is the vector of Riemann constants.

Let us introduce normalized Abelian differentials $d\Omega^{(x)}$ and
$d\Omega^{(t)}$ of the second kind. They are holomorphic on
$\Gamma$ except at the punctures $P_{\pm}$. In the neighborhoods of $P_{\pm}$
they have the form
$$
d\Omega^{(x)}=d(z_{\pm}^{-1}+O(1)),\ \ d\Omega^{(t)}=d(\pm z_{\pm}^{-1}+O(1)).
$$
Normalized means that they have zero $a$-periods.
The vectors of $b$-periods of these differentials are denoted by $2\pi iV$ and
$2\pi i W$, i.e. the coordinates of the vectors $V$ and $W$ are equal to
\beq \label{VW}
V_k={1\over 2\pi i}\oint_{b_k}d\Omega^{(x)},\ \
W_k={1\over 2\pi i}\oint_{b_k}d\Omega^{(t)} .
\eeq
Let $d\Omega^{(n)}$ be a normalized abelian differential of the third
kind with simple poles at the punctures $P_{\pm}$ with residues $\mp 1$.
From the Riemann bilinear relations it follows that the vector of
its $b$-periods satisfies the relation
\beq\label{U}
U_k={1\over 2\pi i}\oint_{b_k}d\Omega^{(n)}=A(P_-)-A(P_+).
\eeq
If we choose a branch of the Abelian integral $\Omega^{(n)}$ near $P_+$ such that
$\Omega^{(n)}=-\ln z_++O(z_+)$, then near $P_-$ it has the form
$$\Omega^{(n)}=\ln z_-+I_0+O(z_-).$$
\begin{th} (\cite{kr3}) The Baker-Akhiezer function is equal to
\beq\label{bak}
\psi_n(x,t,Q)={\theta(A(Q)+nU+xV+tW+Z) \theta(Z)\over
\theta(nU+xV+tW+Z)\theta(A(Q)+Z)}\exp \left(n\Omega^{(n)}+
x\Omega^{(x)}+t\Omega^{(t)}\right).
\eeq
The function $\f_n(x,t)$ given by the formula
\beq\label{st}
\f_n(x,t)=nI_0+\ln{\theta((n+1)U+xV+tW+Z)\over \theta(nU+xV+tW+Z)}
\eeq
is a solution of 2D Toda lattice.
\end{th}
For a generic set of the algebro-geometric data the function
$\f_n(x,t)$ given by (\ref{st}) is a quasi-periodic
meromorphic  function of all the variables $(n,x,t)$. In \cite{zab} the solutions
of 2D Toda lattice which are {\it elliptic in the discrete variable} $n$
were considered. It was found that dynamics of its poles coincides
with the elliptic Ruijsenaars-Schneider system \cite{ruij}. In this paper
we consider solutions that are elliptic in the variable $x$ and
are periodic in $n$.

The condition that $\f_n$ is elliptic in one of the variables is
equivalent to the property that the complex linear subspace in $J(\G)$
spanned by the corresponding directional vector is {\it compact}, i.e.
it is an elliptic curve $\G_0$. In the case of $x$-variable it means
that the vectors $2\omega_{\a}V, \alpha=1,2,$ belong to the lattice ${\cal B}$
defined by (\ref{39}):
\beq\label{100}
2\omega_{\a}V=\sum_k n_k^{\a}e_k+m_k^{\a}B_k,\ \ n_k^{\a},m_k^{\a}\in Z.
\eeq
Here and below $\omega_{1}=\omega, \omega_2=\omega'$ are half-periods
of the elliptic curve $\G_0$.
\begin{th} Let $\G$ be a smooth curve defined by equation (\ref{fam1}) and
let $P_{\pm}$ be preimages on $\G$ of the point $z=0\in \G_0$
with local coordinates in their neighborhoods defined by
the local coordinate $z$ on $\G_0$. Then the corresponding algebro-geometric
solutions given by formula (\ref{st}) satisfy the relation
\beq\label{per}
\f_{n+N}(x,t)=\f_n(x,t)+2N\ln \Lambda,
\eeq
and have the form (\ref{f}), i.e.
$$
\f_n(x,t)=\a_n(t)+\ln {\s(x-x_{n+1}(t)+a)\s(x+x_{n+1}(t)+a)\over
\s(x-x_{n}(t)+a)\s(x+x_{n}(t)+a)}.
$$
The functions $x_{n}(t)$ defined by this representation enjoy equations
(\ref{sys}). The functions $\a_n$ satisfy the relation
\beq\label{con}
4e^{\a_{n-1}(t)-\a_{n}(t)}=\left(1-\dot x_n^2(t)\right)
W(x_n,x_{n+1})W(x_n,x_{n-1}),
\eeq
where
\beq\label{W}
W(u,v)={\s(u-v)\s(u+v)\over \s(2u)}.
\eeq
\end{th}
{\it Proof.} The first statement of the theorem is a direct corollary
of the uniqueness of the Baker-Akhiezer function. The projection
$Q=(w,z)\in \G\to w$ defines $w=w(Q)$ as a function on the curve.
This function is holomorphic on $\G$ outside the puncture $P_+$, where
in has the pole of order $N$, $w=z^{-N}(1+O(z)).$ At the point $P_-$ it
has zero of order $N$, $w=\Lambda^{2N} z^N(1+O(z))$. Therefore, we have
the equality
\beq\label{7}
\psi_{n+N}(x,t,Q)=w(Q)\psi_n(x,t,Q),
\eeq
because the functions defined by its left- and right-hand sides have the same
analytical properties.

Let us consider the functions
\beq
T_{\a}(z)=e^{2\z(z)\omega_{\a}-2\eta_{\a}z}, \ \
\eta_{\a}=\z(\omega_{\a}). \label{bl}
\eeq
They are double-periodic and holomorphic on $\G_0$
except at $z=0$. Again, comparison of analytical
properties of the left- and right-hand sides proves the equality
\beq\label{8}
\psi_n(x+2\omega_{\a},t,Q)=T_{\a}(z)\psi_n(x,t,Q) , \ \ Q=(w,z).
\eeq
The function $e^{\f_n}$ is defined as a ratio of the leading coefficients
of an expansions of $\psi_n$ on two sheets of $\G$. Therefore, it does not
change under the shifts $x\to x+2\omega_{\a}$, and consequently, it is
an elliptic function of the variable $x$.
From (\ref{st}) it follows that if we denote roots of the equation
$\theta(nU+xV+tW+Z)=0$ in the fundamental domain of $\G_0$ by $x_n^j(t), \
j=1,\ldots, D$, then
\beq\label{10}
e^{\f_n(x,t)}=e^{\a_n(t)}\prod_{j=1}^D
{\s(x-x^j_{n+1}(t))\over \s(x-x_n^j(t))}\ .
\eeq
Our next step is to show that $e^{\f_n}$ has only two poles and zeros in
$\G_0$.
\begin{lem} The function $\theta(xV+\xi)$ corresponding to a
smooth algebraic curve $\G$ defined by (\ref{fam1}), as a function of
the variable $x$ is an elliptic theta-function of weight 2, i.e. it can be
represented in the form
\beq\label{11}
\theta(xV+\xi)=r(\xi)\s(x-x^1(\xi))\s(x-x^2(\xi)).
\eeq
\end{lem}
{\it Proof.} Let us find the coefficients of expansion
(\ref{100}). The branching points $z_i^{\pm}$ of $\G$ over $\G_0$
are roots of the equations $P_N^{el}(z)=\pm \Lambda^N$.
In a generic case, when they are distinct, the curve $\G$ is smooth.
The Riemann-Hurwitz formula $2g-2=\nu$ which connects genus $g$
of branching cover of an elliptic curve with a number $\nu$ of branching points,
implies that $\G$ has genus $N+1$. We choose $a_i,b_i$ cycles on it as follows:

\medskip
\noindent
{\it $a_i,\ i=1,\ldots,N-1, $ are cycles around cuts between branching points
$z_i^+,z_i^-$ and $a_{N}$ and $a_{N+1}$ are two preimages of $a$-cycle
on $\G_0$.}
\noindent(We assume that $a$ and $b$-cycles on $\G_0$ correspond to the periods
$2\omega$ and $2\omega'$, respectively.)

From the definition of the differential $d\Omega^{(x)}$ it follows that
\beq\label{12}
d\Omega^{(x)}=d\left(\z(z)-{\eta\over \omega}z\right).
\eeq
Therefore, the coordinates of the vector $V$ defined by
(\ref{VW}) are equal to
\beq\label{V1}
V_i=0,\ i=1,\ldots,N-1,\ \
V_N=V_{N+1}={1\over \pi i}\left(\eta'-{\eta\over \omega} \omega'\right)=
-{1\over 2\omega}.
\eeq
Comparing the vector of $b$-periods of $d\Omega^{(x)}$ with the vector
$(0,\ldots,0,2\omega',2\omega')$ of
$b$-periods of the differential $dz$, considered as a differential on
$\G$, we get
\beq
\oint_{b_{i}}d\Omega^{(x)} =-{\pi i\over 2\omega \omega'}
\oint_{b_{i}}dz, \ i=1,\ldots,N+1. \label{13}
\eeq
The $a$-periods of $dz$ are equal $(0,\ldots,0,2\omega,2\omega)$. Therefore,
$$dz=2\omega (d\Omega_{N}^h+d\Omega_{N+1}^h),$$
where $d\Omega_i^h$ are normalized holomorphic differentials.
From (\ref{13}) we finally obtain that
\beq \label{V2}
2\omega' V=-B_{N}-B_{N+1},
\eeq
where $B_i$ are the vector of $b$-periods
of $d\Omega_i^h$. The monodromy properties of $\theta$-function imply
\beq\label{20}
\theta((x+2\omega)V+Z)=\theta(xV+Z),\ \ \theta((x+2\omega')V+Z)=e^{l(x)}
\theta(xV+Z),
\eeq
where
$$
l(x)=\pi i\left(2x(V_N+V_{N+1})+
B_{N+1,N+1}+B_{N,N}-B_{N,N+1}-B_{N+1,N}+2Z_{N+1}+2Z_N\right)
$$
Using (\ref{V1}) we obtain
\beq \label{21}
dl(x)=-{2\pi i\over \omega}dx.
\eeq
The number $D$ of zeros of the function
$\theta(xV+\xi)$ in the fundamental domain
can be found by integrating of the logarithmic
derivative of this function over the boundary of the domain. From (\ref{20})
and (\ref{21}) it follows that
\beq
D={1\over 2\pi i}\oint_{\p G_{0}} d\ln \theta(xV+Z)= 2.
\eeq
The equality (\ref{11}) is proved. It implies that the
index $j$ in (\ref{10}) takes values $j=1,2$. The sums of zeros and
poles of an elliptic function are equal to each other (modulo periods
of $\G_0$). Hence, $x_n^j(t)$ can be represented
in the form
\beq\label{at}
x_n^1=x_n(t)+a(t),\ \ x_n^2(t)=-x_n(t)+a(t).
\eeq
In order to complete a proof of (\ref{f}) we need only to show that $a(t)$
does not depend on $t$.

Let us substitute (\ref{10}) into (\ref{toda}). A'priory the difference
of the left- and right-hand sides of (\ref{toda})
 is an elliptic function of $x$ with poles of degree 2 at the points
$x_n^j(t)$ and $x_{n+1}^j(t)$. Vanishing of the pole of degree 2 at
$x_n^i$ implies that
\beq\label{30}
\left(\dot x_{n}^i\right)^2-1=F_n^i(x_n^i),\eeq
where
\beq\label{31}
F_n^i(x)=r_n\ {\prod_j\s(x-x_{n+1}^j)\s(x-x_{n-1}^j)
\over \prod_{j\neq i}\s^2(x-x_n^j)}\ ,\ \ r_n=-4e^{\a_n-\a_{n-1}}\ .
\eeq
Vanishing of the pole of degree 1 at $x_n^i$ implies
\begin{eqnarray} \label{32}
\ddot x_n^i&=&\p_x F_n^i(x_n^i)=F_n^i(x_n^i)\left(\p_x \ln F_n^i(x_n^i)\right)=
\nonumber\\
&{=}&\left(\left(\dot x_n^i\right)^2-1\right)
\left(\sum_j \zeta(x_n^i-x_{n+1}^j)+
\zeta(x_n^i-x_{n-1}^j) -2\sum_{j\neq i}\z(x_n^i-x_n^j)\right).
\end{eqnarray}
Substitution of (\ref{at}) in (\ref{31}) shows that
$$F_n^1(x_n^1)=F_n^2(x_n^2)=r_n(t) W(x_n,x_{n+1})W(x_n,x_{n-1}).$$
Hence, we obtain the equality $(\dot x_n^1)^2=(\dot x_n^2)^2,$ which
implies that $\dot a=0$. Equalities (\ref{f}) and (\ref{con}) are proved.
At the same time substitution of (\ref{at}) into (\ref{32}) gives
equations (\ref{sys}).

\section{Generating problem and Lax representation}

In this section we construct the Lax representation for
(\ref{sys}) following an approach proposed in \cite{kr5}, and developed
in \cite{zab,bab,wig} (see their summary in \cite{kr4})). According to
this approach pole dynamics can be obtained simultaneously with its Lax
representation from a specific inverse problem for a {\it linear} operator
with elliptic coefficients.

In the most general form this inverse problem is to
find linear operators with elliptic coefficients that have
{\it sufficiently enough} double-Bloch solutions.
A {\it meromorphic} function $f(x)$ is called double-Bloch if
it has the following monodromy properties:
\beq
f(x+2\omega_{\alpha})=B_{\alpha} f(x), \ \  \alpha=1,2.\label{g1}
\eeq
The complex numbers $B_{\alpha}$ are  called
{\it Bloch multipliers}.  (In other words, $f$ is a meromorphic section of a
vector bundle over the elliptic curve.)
It turns out that existence of the double-Bloch solutions is so
restrictive that only in exceptional cases such solutions do exist.

The basis in the space of double-Bloch functions can be written in terms
of the fundamental function $\Phi(x,z)$ defined by the formula
\beq
\Phi(x,z)={\sigma(z-x)\over \sigma(z) \sigma(x)}\
e^{\zeta(z)x}. \label{phi}
\eeq
From the monodromy properties of the Weierstrass functions
it follows that $\Phi$, considered
as a function of $z$, is double-periodic:
$\Phi(x,z+2\omega_{\alpha})=\Phi(x,z) ,$
though it is not elliptic in the classical sense due to
essential singularity at $z=0$ for $x\neq 0$.
As a function of $x$, the function $\Phi(x,z)$ is double-Bloch function, i.e.
$$
\Phi(x+2\omega_{\alpha}, z)=T_{\alpha}(z) \Phi (x, z),
$$
where $T_{\a}(z)$ are given by (\ref{bl}).
In the fundamental domain of the lattice defined by
$2\omega_{\alpha}$ the function $\Phi(x,z)$ has a unique pole at the point
$x=0$:
\beq
\Phi(x,z)=x^{-1}+O(x). \label{j}
\eeq
Let $f(x)$ be double-Bloch function with simple
poles $x_i$ in the fundamental domain and with Bloch multipliers
$B_{\alpha}$ (such that at least one of them is not equal to $1$). Then it
can be represented in the form:
\beq
f(x)=\sum_{i=1}^N c_i\Phi(x-x_i,z) e^{k x},\label{g2}
\eeq
where $c_i$ is the residue of $f$ at $x_i$ and $(z,k)$ are parameters
such that
$
B_{\alpha}=T_{\alpha}(z)\exp(2\omega_{\alpha}k).
$

Now we are in position to present the generating problem for (\ref{sys}).

\begin{th} The equation
\beq\label{e0}
(\p_t+\p_x)\Psi_{n}=2\Psi_{n+1}+v_n(x,t)\Psi_n
\eeq
with an elliptic coefficient of the form
\beq\label{v}
v_n(x,t)=\g_n(t)+
\sum_{i=1}^2 \left[h_n^i(t)\z(x-x_{n}^i(t))-h_{n+1}^i(t)\z(x-x_{n+1}^i(t))
\right],
\eeq
where
\beq\label{50}
x_n^1(t)=x_n(t)+a,\ \ x_n^2(t)=-x_n(t)+a,\ \ a=const,
\eeq
has two linear independent double-Bloch solutions with Bloch multipliers
$T_{\a}(z)$ (for some $z$), i.e. solutions of the form
\beq\label{ans}
\Psi_n(x,t)=\sum_{i=1}^2 c_n^i\Phi(x-x_n^i(t),z)
\eeq
if and only if the functions $x_n(t)$ satisfy equation (\ref{sys}).

If equation (\ref{e0}) has two linear independent solutions of the form
(\ref{ans}) for some $z$, then they exist for all values of $z$.
\end{th}
{\it Proof.} Let us substitute (\ref{ans}) into  (\ref{e0}). The both sides
of the equation are double-Bloch functions with the same Bloch multipliers
and with the pole of order 2 at $x_n^i$, and the simple pole at $x_{n+1}^i$.
They coincide iff the coefficients of their singular parts at these points
are equal to each other.
The equality of the coefficients at $(x-x_n^i)^{-2}$ implies
\beq
h_n^i=\dot x_n^i-1 .\label{s2}
\eeq
The equality of residues at $x_{n+1}^i$ is equivalent to
the equation
\beq
c_{n+1}^i=2^{-1}h_{n+1}^i\sum_j\Phi(x_{n+1}^i-x_n^j) c_n^j. \label{laxL}
\eeq
The equality of residues at $x_{n}^i$ is equivalent to
the equation
\beq
\p_tc_{n}^i=M_n^ic_n^i+h_{n}^i\sum_{j\neq i}\Phi(x_{n+1}^i-x_n^j) c_n^j, \label{laxM}
\eeq
where
\beq\label{mi}
M_n^i=\g_n-\sum_j h_{n+1}^j\z(x_n^i-x_{n+1}^j)+
\sum_{j\neq i}h_n^j\z(x_n^i-x_n^j).
\eeq
Equations (\ref{laxL}) and (\ref{laxM}) are linear equations for $c_n^i$.
Their compatibility is just a system of the equations:
$$
\p_t(\ln h_{n+1}^i)\Phi(x_{n+1}^i-x_n^j)+
(\dot x_{n+1}^i-\dot x_n^j)\Phi'(x_{n+1}^i-x_n^j)
=(M_{n+1}^i-M_n^j)\Phi(x_{n+1}^i-x_n^j)+
$$
\beq
+\sum_{k\neq i}^{}\Phi(x_{n+1}^i-x_{n+1}^k)h^k_{n+1}\Phi(x_{n+1}^k-x_n^j)
-\sum_{k\neq j}^{}\Phi(x_{n+1}^i-x_{n}^k)h^k_{n}\Phi(x_{n}^k-x_n^j),
\label{LM1}
\eeq
which can be written in the matrix form:
\beq\label{LM}
\p_t L_n=M_{n+1}L_n-L_nM_n,
\eeq
where $L_n$ and $M_n$ are matrices defined by the right-hand sides of
(\ref{laxL}) and (\ref{laxM}). Equations (\ref{LM}) are necessary and
sufficient conditions for the existence of solutions of (\ref{e0}) which
have the form (\ref{ans}). Therefore, the following statement
completes a proof of the theorem.

\begin{lem} Let $L_n=\left(L_n^{ij}(t,z)\right)$ and
$M_n=\left(M^{ij}_n(t,z)\right)$ be defined by the formulae
\beq\label{LL}
L_n^{ij}=2^{-1}h_{n+1}^i \Phi(x_{n+1}^i-x_n^j,z),\ \ M_n^{ii}= M_n^i,\ \
M_{ij}=h_n^i\Phi(x_n^i-x_n^j,z),\ i\neq j,
\eeq
where $x_n^1=x_n,\ x_n^2=-x_n$, $h_n^i=\dot x_n^i-1$, and $M_n^i$ is
given by (\ref{mi}) with $\g_n$ such that
\beq\label{e12}
\g_n-\g_{n-1}=d_t\ln\left({(\dot x_n^2-1)\s^2(2x_n)\over
\s(x_n-x_{n+1})\s(x_n+x_{n+1})\s(x_n-x_{n-1})\s(x_n+x_{n-1})}\right).
\eeq
Then they satisfy equation (\ref{LM}) if and only if the functions $x_n(t)$
solve equations (\ref{sys}).
\end{lem}
Note, that (\ref{e12}) defines $\g_n(t)$ up to a constant shift
$\g_n(t)\to \g_n(t)+g(t)$, which corresponds to the gauge
transformation $\Psi_n\to e^g\Psi_n$ of equation (\ref{e0}), and
which does not effect equations for $x_n$.

\medskip

\noindent{\it Proof.}
The right- and left-hand sides of (\ref{LM1})
are double-periodic functions of $z$ that are holomorphic except at $z=0$,
where they have the form  $O(z^{-2})\exp((x_{n+1}^i-x_n^j)\z(z))$.
Such functions are equal if and only if the corresponding
coefficients at $z^{-2}$ and $z^{-1}$ are equal.
The equality of the coefficients at $z^{-2}$ gives
\beq\label{e1}
(\dot x_{n+1}^i-\dot x_{n}^j)=h_{n+1}^i-h_n^j+\sum_{k} (h_n^k-h_{n+1}^k)=
h_{n+1}^i-h_n^j,
\eeq
which is fulfilled due to (\ref{s2}) (the second equality in (\ref{e1})
holds because $v(x,t)$ is an elliptic function
of $x$ and, therefore, a sum of its residues is equal to zero).

The equality of the coefficients at $z^{-1}$ in the expansion
of (\ref{LM1}) at $z=0$ gives
$$
\p_t(\ln h_{n+1}^i)-(\dot x_{n+1}^i-\dot x_n^j)\z(x_{n+1}^i-x_n^j)=M_{n+1}^i-
M_{n}^j+
$$
\beq\label{e2}
+\sum_{k\neq i}h_{n+1}^k \left[\z(x_{n+1}^i-x_{n+1}^k)+\z(x_{n+1}^k-x_n^j)
\right]-
\sum_{k\neq j}h_{n}^k \left[\z(x_{n+1}^i-x_{n}^k)+\z(x_{n}^k-x_n^j)\right].
\eeq
The second line in (\ref{e2}) is equal up to the sign to the sum
of residues at $x_n^k,\ k\neq j,$ and at $x_{n+1}^k,\ k\neq i,$ of the
elliptic function
$$\widetilde v_n(x,t)=v_n(x,t)\left[\zeta(x_{n+1}^{i}-x)+\z(x-x_{n}^j)\right].$$
Therefore, it equals to the sum of residues of this function at $x_{n+1}^i$
and $x_n^j$. We have
$$
\res_{x_{n+1}^i}\widetilde v_n(x,t)+\res_{x_n^j}\widetilde v_n(x,t)=
(h_n^j-h_{n+1}^i)\z(x_{n+1}^i-x_n^j)+M_n^j-
$$
\beq\label{r}
-\g_n-\sum_kh_n^k\z(x_{n+1}^i-x_n^k)+
\sum_{k\neq i}h_{n+1}^k\z(x_{n+1}^i-x_{n+1}^k)\ .
\eeq
Substitution of the right-hand side of the last equality into
(\ref{e2}) implies (after the shift $n+1\to n$)
\beq\label{e3}
{\dot h_n^i\over h_n^i}=\g_{n}-\g_{n-1}+\sum_{k\neq i}2h_n^k\z(x_{n}^i-x_n^k)
-\sum_{k}\left[h_{n+1}^k\z(x_{n}^i-x_{n+1}^k)+
h_{n-1}^k\z(x_{n}^i-x_{n-1}^k)\right].
\eeq
From (\ref{s2}) it follows, that (\ref{e3}) can be rewritten in the form
\beq\label{e4}
{\ddot x_n^i\over \dot x_n^i-1}=\p_xG_n^i(x_n^i)+\p_tG^i_n(x_n^i),
\eeq
where  the function
\beq\label{e5}
G_n^i(x)=a_n+
\ln\left({\prod_{k}\s(x-x_{n+1}^k)\s(x-x_{n-1}^k)\over
\prod_{k\neq i}\s^2(x-x_n^k)}\right) ,\ \ \  \p_ta_n=\g_n-\g_{n-1},
\eeq
depends on $t$ through the dependence on $t$ of $x_m^i$ and $a_n$, only.
By chain rule we have
\beq\label{e6}
{d}_t\left(G_n^i(x_n^i)\right)=\dot x_n^i\p_xG_n^i(x_n^i)+
\p_tG_n^i(x_n^i),\ \ {d}_t={d\over dt}.
\eeq
Therefore,
\beq\label{e7}
{\ddot x_n^i\over \dot x_n^i-1}=(1-\dot x_n^i)\p_xG_n^i(x_n^i)+d_t
\left(G^i_n(x_n^i)\right),
\eeq
From (\ref{50}) it follows that
$$
G^1_n(x_n^1)=G^2_n(x_n^2)=G_n(x_n),\ \
\p_xG^1_n(x_n^1)=-\p_xG^2_n(x_n^2)=\p_xG_n(x_n),
$$
where
\beq\label{e8}
G_n(x)=a_n+
\ln\left({\s(x-x_{n+1})\s(x+x_{n+1})\s(x-x_{n-1})\s(x+x_{n-1})\over
\s^2(x+x_n)}\right).
\eeq
Therefore, equations (\ref{e7}) for $i=1,2,$ have the form
\begin{eqnarray}
{\ddot x_n\over \dot x_n-1}&=&d_t\left(G_n(x_n)\right)-
(\dot x_n-1)\p_xG_n(x_n), \label{e9} \\
{\ddot x_n\over \dot x_n+1}&=&d_t\left(G_n(x_n)\right)-
(\dot x_n+1))\p_xG_n(x_n). \label{e10}
\end{eqnarray}
Equations (\ref{e9}) and (\ref{e10}) are equivalent to the equations:
\beq\label{e11}
\ddot x_n=(\dot x_n^2-1)\p_xG_n(x_n),\ \
d_t\left(G_n(x_n)\right)=d_t\ln{(\dot x_n^2-1)} .
\eeq
The first among them, coincides with equations (\ref{sys}) for $x_n$,
and the second one (compare it with (\ref{con})) is equivalent to the definition
of $\g_n$ by (\ref{e12}). Lemma and Theorem are proved.

\section{Direct problem. Spectral curves.}

In this section we consider periodic in $n$
solutions of equations (\ref{sys}).
\begin{lem}
Let $x_n(t)=x_{n+N}(t)$ be a solution of (\ref{sys}).
Then
\beq
I=\prod_{n=1}^N\left(
{\s(x_n-x_{n+1})\s(x_n+x_{n+1})\s(x_n-x_{n-1})\s(x_n+x_{n-1})\over
(\dot x_n^2-1)\s^2(2x_n)}\right),
\label{02}
\eeq
is an integral of motion,\ $I=const$, and the monodromy matrix
\beq
T(t,z)=\prod_{n=0}^{N-1}L_n(t,z) \label{mon}
\eeq
satisfies the Lax equation
\beq
\p_t T=[M_0, T].    \label{m0}
\eeq
\end{lem}
{\it Proof.} If $x_n(t)$ is periodic in $n$, then the corresponding
matrix functions $L_n(t,z)$ and $M_n(t,z)$ defined by (\ref{LL})
satisfy the relations
\beq
L_{n+N}=L_n ,\ \ M_{n+N}=M_n-d_t(\ln I).\label{01}
\eeq
Therefore, equation (\ref{LM}),
$
\p_t L_n=M_{n+1}L_n-L_nM_n,
$
implies that
\beq\label{03}
\p_t T=-d_t(\ln I)\ T+[M_0, T].
\eeq
Note, that if $\p_t I=0$ then (\ref{03}) coincides with (\ref{m0}), and
therefore, the second statement of the Lemma follows from the first one.

Equation (\ref{03}) implies that the function
\beq
P(z)=I(t)({\rm tr} \ T(t,z))
\eeq
is {\it time-independent}.

Matrix entries of $L_n$ are double-periodic functions that are holomorphic
on $\G_0$ except at $z=0$. Therefore, $({\rm tr}\ T)$ is also double-periodic
and holomorphic on $\G_0$ outside $z=0$. In order to prove that
this function is meromorphic on $\G_0$, it is enough to note
that $L_n$ has the form
\beq
L_n(t,z)=g_{n+1}\widetilde L_n g_n^{-1},\label{3.5}
\eeq
where
$$g_n=\left(\begin{array}{cc} e^{x_n\z(z)} & 0\\
                 0 & e^{-x_n\z(z)} \end{array}\right)\ .
$$
From (\ref{phi}) it follows that in the neighborhood of $z=0$
\beq\label{3.6}
\widetilde L_n=(z)^{-1}\widetilde L_n^0+\widetilde L_n^1+O(z),
\eeq
where
\beq\label{3.61}
\widetilde L_n^0\ =\ {1\over 2}\matrix{1-\dot x_{n+1} & 1-\dot x_{n+1}\\
                                       1+\dot x_{n+1} & 1+\dot x_{n+1}}
\eeq
and
\beq\label{3.62}
\widetilde L_n^1\ =\ {1\over 2}
\left(\begin{array}{cc} 1-\dot x_{n+1} & 0\\ 0 & 1+\dot x_{n+1}
\end{array}\right)
\left(\begin{array}{cc} -\z(x_{n+1}-x_n) & -\z(x_{n+1}+x_n) \\
       \  \z(x_{n+1}+x_n) & \ \z(x_{n+1}-x_n)\end{array}\right) .
\eeq
Therefore,
\beq
{\rm tr} \ T={\rm tr}\left( \prod_{n=0}^{N-1}\widetilde L_n(t,z)\right)=
z^{-N}(1+0(z)). \label{3.7}
\eeq
The last equality shows that $({\rm tr}\ T)$ is a monic elliptic
polynomial $P_N^{el}(z)$. Therefore,
at $z=0$ we have $P(z)=I(t)z^{-N}(1+0(z))$.
Hence, $I(t)$ is an integral of (\ref{sys}) because $P(z)$ does
not depend on $z$. Lemma is proved.

Due to (\ref{m0}) the spectral curve $\G$ defined by the characteristic
equation
\beq
R(w,z)\equiv \det\left(w-T(t,z)\right)=w^2-({\rm tr}\ T)w+\det T=0
\label{spec}
\eeq
is {\it time-independent}.
\begin{lem}
The characteristic equation (\ref{spec}) has the form (\ref{fam}) .
\end{lem}
{\it Proof.}
We have already proved that $({\rm tr} \ T)$ has the form (\ref{el}).
The relation $\Phi(x,z)\Phi(-x,z)=\wp(z)-\wp(x)$,
which is equivalent to the addition formula for the Weierstrass $\s$-function,
implies
\beq
\det L_n(t,z)=
2^{-2} (\dot x_{n+1}^2-1)\left[\wp(x_{n+1}-x_n)-\wp(x_{n+1}+x_n)\right]\ .
\label{3.3}
\eeq
Therefore, though $L_n(t,z)$ depends on $z$, its  determinant
does not depend on $z$. Hence, $(\det \ T)$
is also $z$-independent. As it does not depend on $t$, we
identify $\Lambda^{2N}$ in (\ref{fam1}) with
\beq
\Lambda^{2N}=\det T(t,z)=2^{-2N} \prod_{n=0}^{N-1}
(\dot x_n^2-1)\left(\wp(x_{n}-x_{n-1})-\wp(x_{n}+x_{n-1})\right)=2^{-2N}e^{H},
\label{3.4}
\eeq
where $H$ is the Hamiltonian of system (\ref{sys}).
Lemma is proved.

For a generic point $Q$ of the spectral curve $\G$, i.e. for a pair
$(w,z)$ that satisfies (\ref{spec}) there exists a unique
solution $C_n=(c_n^i(t,Q))$ of the equations
\beq
C_{n+1}(t,Q)=L_n(t,z)C_n(t,Q), \ \ \p_tC_n(t,Q)=M_n(t,z),
\label{3.1}
\eeq
such that
\beq
C_{n+N}(t,Q)=w C_n(t,Q), \label{3.2}
\eeq
and normalized by the condition
\beq\label{norm}
c_0^1(0,Q)\Phi(-x_0(0),z)+c_0^2(0,Q)\Phi(x_0(0),z)=1. \label{3.8}
\eeq
{\it Remark.} Normalization (\ref{3.8}) corresponds to a usual normalization
$\Psi_0(0,0,Q)=1$ of the solution $\Psi_n(x,t,Q)$ of (\ref{e0}) defined
by (\ref{ans}).
\begin{th}
The coordinates $c_n^i(t,Q)$ of the vector-valued function $C_n(t,Q)$
are meromorphic functions on $\G$ except at
the preimages $P_{\pm}$ of $z=0$. Their poles $\g_1,\ldots,\g_{N+1}$
do not depend on $n$ and $t$. The projections $z(\g_s)$ of these poles on
$\G_0$ satisfy the constraint
\beq \label{cons}
\sum_{s=1}^{N+1} z(\g_s)=0.
\eeq
In the neighborhoods of $P_{\pm}$ the coordinates of $C_n(t,Q)$ have the
form
\begin{eqnarray}
c_n^1(t,Q)&=&z^{\mp n}\chi_{n,\pm}^{1}(t,z)e^{(\pm t+x_n(t))z^{-1}},
\label{3.9}\\
c_n^2(t,Q)&=&z^{\mp n}\chi_{n,\pm}^{2}(t,z)e^{(\pm t-x_n(t))z^{-1}}, \label{3.10}
\end{eqnarray}
where $\chi_{n,\pm}^{i}(t,z)$ are regular functions of $z$:
\beq
\chi_{n,+}^{i}(t,z)=z\chi_{n,+}^{i}(t)+O(z^2),\ \
\chi_{n,-}^{i}(t,z)=\chi_{n,-}^{i}(t)+z\chi_{n,-}^{i,1}(t)+O(z^2),\label{3.12}
\eeq
such that the leading coefficients of their expansions have the form:
\beq\label{3.130}
\chi^{1}_{n,+}(t)=c(t)(1-\dot x_n),\ \chi_{n,+}^{2}(t)=c(t)(1+\dot x_n),\ \
c(0)=1,
\eeq
\beq
\chi^{1}_{n,-}(t)=s_n(t),\ \chi_{n,-}^{2}(t)=-s_n(t), \label{3.13}
\eeq
where functions $s_n$ satisfy the relation
\beq\label{3.131}
s_{n+1}=
2^{-2}(\dot x_{n+1}^2-1)\left[\wp(x_{n+1}-x_n)-\wp(x_{n+1}+x_n)\right]\ s_n.
\eeq
\end{th}
{\it Proof.} Vector-columns $S_{n}^{(1)}$ and $S_n^{(2)}$ of
the matrix-function
\beq\label{3.14}
S_0^{ij}=\delta_{ij},\ \ S_n(t,z)=\prod_{m=0}^{n-1}L_m(t,z),\ n>0,
\eeq
are holomorphic functions on $\G_0$ except at $z=0$. They satisfy the
equation $S_{n+1}^{(i)}=L_nS_{n}^{(i)}$. Therefore,
the Bloch solution $C_n$ of (\ref{3.1}) has the form
\beq
C_n(t,Q)=h_1(Q)S_n^{(1)}(t,z)+h_2(Q)S_n^{(2)}(t,z), \label{3.15}
\eeq
where $h_i(Q), \ Q=(w,z)\in \G,$ are the coordinates of the normalized
eigenvector of the monodromy matrix $T(z)$,
corresponding to the eigenvalue $w$. They are equal to
\beq
h_1(Q)={1\over r(Q)}T^{12}(z),\ \ h_2(Q)={1\over r(Q)}(w-T^{11}(z)), \label{3.16}
\eeq
where $T^{ij}(z)$ are entries of the monodromy matrix and the normalization
constant $r(Q)$ equals
\beq
r(Q)=T^{12}(z)\Phi(-x_0(0),z)+\left(w-T^{11}(z)\right)\Phi(x_0(0),z).
\label{3.17}
\eeq
The function $r(Q)$ has the pole of degree $N+1$ at $P_+$ and the
pole of degree $N$ at $P_-$. Therefore, it has $2N+1$ zeros.

Let us show that $N$ of these zeros are situated over roots of the equation
$T^{12}(z)=0$ on one of the sheets of $\G$. Indeed,
if $T^{12}(z)=0$, then eigenvalues $w(z)$ of the monodromy matrix
are equal to $T^{11}(z)$ or $T^{12}(z)$. Therefore, $r=0$ at the points
$Q=(T^{11}(z),z)$. Equations (\ref{3.16}) imply that $C_n$ has no
poles at these points. The poles $\g_s$ of $C_n(t,Q)$ on $\G$ outside
the punctures $P_{\pm}$ are the other zeros of $r(Q)$ and do not
depend on $n$ and $t$. Let us prove now that they satisfy (\ref{cons}).

The function $r^*(z)=r(Q)r(Q^{\s})$ with $\s:Q\to Q^{\s}$ as permutation
of sheets of $\G$, is a well-defined function on $\G_0$ with the pole of
degree $2N+1$ at $z=0$. As it was shown above it is divisible by $T^{12}(z)$.
Therefore, the ratio $r^*(z)/T^{12}(z)$
is an elliptic function with the pole of degree $N+1$ at $z=0$ and zeros at
the points $z(\g_s)$. Divisors of zeros and poles of an elliptic function
are equivalent. Therefore, (\ref{cons}) is proved.

From (\ref{3.5}) it follows that the vector-function
$\widetilde C_n=g_{n}^{-1}C_n$ is a Bloch solution of the equation
$\widetilde C_{n+1}=\widetilde L_n \widetilde C_n.$
Let us first consider the neighborhood of the puncture $P_+$, which
corresponds to the branch $w=z^{-N}(1+O(z))$
of the eigenvalue of the monodromy matrix.

The vector-function $X_{n}(t)$ with the coordinates given by (\ref{3.130})
satisfies the equation $X_{n+1}=\widetilde L_n^0X_{n}$, where
$\widetilde L_n^0$ is defined in (\ref{3.6}).
That implies that in the neighborhood of $P_+$ the vector-function
$C_n(t,Q)$ has the form stated in the theorem up to a time-dependent
factor $f_+(t,z)$. Substitution of (\ref{3.9},\ref{3.10}) into
the equation $\p_t C_n=M_n C_n$ shows that $\p_t f=O(z)$. Therefore,
the analytical properties of $C_n$ near $P_+$ are established.

Now we are going to prove by induction that at $P_-$ equalities
(\ref{3.9}, \ref{3.10}), and (\ref{3.13}) hold. For $n=0$ they are fulfilled
by the normalization conditions. Let us prove first, that if
(\ref{3.9}, \ref{3.10}), and (\ref{3.13}) hold for $n'\leq n$, then
\beq\label{3.171}
2\kappa_n=\left(\z(x_{n+1}+x_n)-\z(x_{n+1}-x_n)\right)s_{n}+
\chi_{n,-}^{1,1}+\chi_{n,-}^{1,2}=0.
\eeq
Indeed, the equation $C_{n+1}=L_nC_n$ implies that
$\widetilde C_{n+1}$ at $P_-$ has the form
\beq\label{3.170}
\widetilde C_{n+1}=z^{n}\left(\begin{array}{c} (1-\dot x_{n+1})\kappa_n\\
(1+\dot x_{n+1})\kappa_n \end{array}\right)+O(z^{n+1}).
\eeq
Hence,
\beq
\widetilde C_N=\left(\prod_{m=n+1}^{N-1}L_m\right)\widetilde C_{n+1}=
z^{2n-N-1}\left(\begin{array}{c} (1-\dot x_{0})\kappa_n\\
(1+\dot x_0)\kappa_n \end{array}\right)+
O\left(z^{2n-N}\right).
\eeq
If $\kappa_n\neq 0$, then the last equality contradicts
the monodromy property $\widetilde C_N=w\widetilde C_0=O(z^{N})$.
Therefore, $\kappa_n=0$, and then (\ref{3.170}) shows that
$\widetilde C_{n+1}$ has zero of order $n+1$ at $P_-$. Therefore, a step
of induction for equalities (\ref{3.9}, \ref{3.10}) is proved.
The same arguments show that if
(\ref{3.13}) does not hold then the vector $\widetilde C_N$ has
zero of order $(2n-N)$ which again contradicts the relation $\widetilde C_N=
O(z^N).$

Equalities (\ref{3.9}, \ref{3.10}) near $P_-$ are proved, possibly
up to a time-dependent factor $f_-(t,z)$. Their substitution into
the equation $\p_tC_n=M_nC_n$ shows that $\p_t f_-=O(z)$ and
completes a proof of (\ref{3.9}, \ref{3.10}),  and (\ref{3.13}).

Let ${\cal C}_n(z)$ be a matrix formed by the vectors $C_n(t,Q_i(z))$,
corresponding to two different sheets $Q_i(z)=(w_i(z),z)$ of $\G$.
This matrix is defined up to permutation of sheets.
From (\ref{3.9}-\ref{3.13}) it follows that in the neighborhood of $z=0$
\beq\label{X1}
{\cal C}_n(z)=
\left(\begin{array}{cc} e^{x_n\z(z)} & 0\\ 0 & e^{-x_n\z(z)} \end{array}\right)
\left[\left(\begin{array}{cc} (1-\dot x_n)c & s_n \\
(1+\dot x_n)c & -s_n \end{array}\right)
+O(z)\right]
\left(\begin{array}{cc} z^{-n+1}e^{t\z(z)} & 0\\ 0 & z^ne^{-t\z(z)}
\end{array}\right)
\eeq
Therefore,
\beq\label{X2}
\det {\cal C}_n=-2cs_n z+O(z^2),
\eeq
and from the definition of ${\cal C}_n$ we have
${\cal C}_{n+1}=L_n{\cal C}_n$. Hence,
\beq\label{X3}
s_{n+1}=s_n\det L_n,
\eeq
which coincides with (\ref{3.131}). Theorem is proved.

The correspondence which
assigns to each solution $x_n(t)=x_{n+N}(t)$ of (\ref{sys})
a set of algebro-geometric
data $\{\G, D\}$, is a direct spectral transform.
The following statement shows that the results of Section 2, can be seen as
the inverse spectral transform.
\begin{cor} The solution
\beq\label{3.20}
\Psi_n(x,t,Q)=c_n^1(t,Q)\Phi(x-x_n(t),t)+c_n^2(t,Q)\Phi(x+x_n(t),t)
\eeq
of equation (\ref{e0}) is equal to $\Psi_n(x,t,Q)=c(t)\psi_n(x,t,Q)$,
where $\psi_n(x,t,Q)$ is the Baker-Akhiezer function corresponding to $\G$
and the divisor $D$ of the poles of $C_n$; the factor $c(t)$
is defined in (\ref{3.130}).

All the solutions $x_n(t)$ of (\ref{sys}) have the
form $x_n={1\over 2}(x_n^1-x_n^2)$, where $x_n^i(t)$ are roots of the equation
\beq\label{for}
\theta(nU+x_n^i(t)V+Wt+Z)=0.
\eeq
Here $\theta(z)$ is the Riemann theta-function corresponding to $\G$;
vectors $U,V,W$ are defined by (\ref{VW}, \ref{U});
vector $Z$ corresponds to the divisor $D$ via the Abel transform.
\end{cor}
As follows from the Theorem 4.1, the function $\Psi_n$ defined by (\ref{3.20})
has the same analytical properties on $\G$ as the function $c(t)\psi_n$.
Therefore, they coincide. Equation (\ref{for}) immediately follows from
the formula (\ref{bak}) for $\psi_n.$

\section{Action-angle variables}
Until now we have not used the Hamiltonian structure of equations
(\ref{sys}). Moreover, a'priory it is not clear why a system that has
arisen as a pole system of elliptic solutions of the 2D Toda lattice
is Hamiltonian. The general algebro-geometric approach which allows to
derive a Hamiltonian structure starting from the Lax representation
was proposed and developed in \cite{kp1,kp2,kr4}.

The main goal of this section is to construct action-angle variables for
(\ref{sys}). First of all, let us summarize necessary results
of the previous sections. A point $(p_n,x_n)$ of the phase space $\M$
of the system defines a matrix function $L_n(z)$
with the help of the formulae
\beq \label{h1}
L_n^{ij}=2^{-1}h_{n+1}^i \Phi(x_{n+1}^i-x_n^j,z),
\eeq
\beq\label{h2}
x_n^1=x_n,\ x_n^2=-x_n,\ h_n^1=h_n-1,\ h_n^2=-h_n-1,\
h_n={1+e^{p_n}\over 1-e^{p_n}}\ .
\eeq
This function defines the spectral curve $\G$ (with the help of
(\ref{spec})), and the divisor $D$ of poles
$\g_1,\ldots,\g_{N+1}$ of the Baker-Akhiezer function
$C_n(Q)=(c_n^1(Q),c_n^2(Q))$
\beq\label{h3}
C_{n+1}(Q)=L_n(z)C_n(Q),\ \ C_N(Q)=wC_0(Q),\ \ Q=(w,z)\in \G,
\eeq
normalized by the condition
\beq\label{h4}
c_0^1(0,Q)\Phi(-x_0,z)+c_0^2(Q)\Phi(x_0,z)=1. \label{3.8'}
\eeq
The divisor $D$ satisfies (\ref{cons}), i.e. defines a point of an odd part
of the Jacobian $J^{Pr}(\G)\in J(\G)$, which is defined as a fiber of the
projection
\beq\label{h51}
\{\g_1,\ldots,\g_{N+1}\}\in J(\G)\longmapsto
2\omega \phi_+=\sum_{s=1}z(\g_s)\in \G_0,
\eeq
corresponding to $\phi_+=0$. All the fibers are equivalent and can be
identified with the Prym variety of $\G$.
Note, that a shift of $\phi_+$ corresponds to the shift $x\to x+a$ for
the solution (\ref{f}) of 2D Toda lattice.

The correspondence
\beq\label{h5}
(p_n,x_n)\in \M\longmapsto \{\G,D\in J^{Pr}(\G)\}
\eeq
is an isomorphism. The coefficients $(u_i, \Lambda)$ of equation (\ref{fam1})
are integrals of the Hamiltonian system (\ref{sys}).
Equations (\ref{sys}) on a fiber over $\G$ of the map (\ref{h5})
are linearized by the Abel transform (\ref{ab}).

The main goal of this section is to construct the action variables
that are canonically conjugated to the coordinates
$\phi_1,\ldots,\phi_{N-1},\phi_-:$
\beq\label{h60}
\phi_k=\sum_{s=1}^{N+1}A_k(\g_s), \ \phi_-=\phi_N-\phi_{N+1},
\eeq
on the Prymmian $J^{Pr}(\G)$. Note, that $\phi_+=\phi_N+\phi_{N+1}$.
\begin{th} The transformation
\beq \label{h61}
(x_n,p_n)\longmapsto (\phi_1,\ldots,\phi_{N-1},\phi_-;I_1,\ldots I_N)
\eeq
where $I_k$ are $a$-periods of the differential $dS=\ln (\Lambda^{-N}w)dz$:
\beq\label{h62}
I_k=\oint_{a_k}\ln (\Lambda^{-N}w)dz,
\eeq
is a canonical transformation, i.e.
\beq\label{h63}
\sum_{n=1}^Ndp_n\wedge dx_n=
\sum_{k=1}^{N-1}\left(\delta I_k\wedge \delta \phi_k\right)+
\delta I_N\wedge \delta \phi_-.
\eeq
\end{th}
{\it Proof.} First of all, following the approach proposed in
\cite{kp1}, we define a symplectic structure on $\M$ in terms of
the Lax operator and its eigenfunctions. After that we will calculate
it in two different ways which immediately imply (\ref{h63}).

The external differential $\delta L_n(z)$ can be seen as an operator-valued
one-form on $\cal M$. Canonically normalized eigenfunction $C_n(Q)$
of $L_n(z)$ is the vector-valued function on $\cal M$.
Hence, its differential is a vector-valued one-form.
Let us define a two-form $\omega$ on ${\cal M}$
by the formula
\beq
\omega={1\over 2}\left(\res_{P_+} \Omega+\res_{P_-} \Omega\right),
\label{54}
\eeq
where
\beq\label{541}
\Omega=<C^*_{n+1}(Q) \delta L_n(z)\wedge \delta C_n(Q)>dz.
\eeq
Here and below $<\cdot>$ stands for the sum over a period of a periodic in $n$
function, i.e.
$$<f_n>=\sum_{n=0}^{N-1}f_n;$$
$C_n^*(Q)$ is the dual Baker-Akhiezer function, which is defined
as a co-vector (row-vector) solution of the equation
\beq\label{h6}
C_{n+1}^*(Q)L_n(z)=C_n^*(Q), \ \ C_N^*(Q)=w^{-1}C_0^*(Q),
\eeq
normalized by the condition
\beq\label{h7}
C_0^*(Q)C_0(Q)=1.
\eeq
The form $\omega$ can be rewritten as
\beq
\omega={1\over 2}\res_{0} \ {\rm Tr}
<\left(\C_{n+1}^{-1}(z) \delta L_n(z)\wedge \delta  \C_n(z)\right)> dz,
\label{55}
\eeq
where $\C_n(z)$ is a matrix with the columns $C_n(Q_j(z)), \ Q_j(z)=(z,w_j)$
corresponding to different sheets of $\G$.

Note, that $C_n^*(Q)$ are rows of the matrix $\C_n^{-1}(z)$.
That implies that $C_n^*(Q)$ as a function on the spectral curve is:
meromorphic outside the punctures; has poles
at the branching points of the spectral curve, and zeros at the poles $\g_s$ of
$C_n(Q)$. These analytical properties are used in the proof of the following
lemma.
\begin{lem} The two-form $\omega$ equals
\beq
\omega=\sum_{s=1}^{N+1}\delta z(\g_s)\wedge \delta \ln w(\g_s). \label{61}
\eeq
\end{lem}
The meaning of the right-hand side of this formula is as follows.
The spectral curve by definition arises with the meromorphic function $w(Q)$
and multi-valued holomorphic function $z(Q)$.
Their evaluations $w(\g_s),\ z(\g_s)$ at the points
$\g_s$ define functions on $\cal M$, and the wedge product of their
external differentials is a two-form on $\cal M$.

\noindent{\it Proof.}
The differential $\Omega$, defined by (\ref{541}) is a meromorphic
differential on the spectral curve (the essential singularities of the
factors cancel each other at the punctures). Therefore, the sum of its
residues at the punctures is equal to the sum of other residues with negative
sign. There are poles of two types.

First of all, $\Omega$ has poles at the poles $\g_s$ of $C_n$.
Note, that $\delta C_n$ has the pole of the second order
at $\g_s$. Taking into account that $C_n^*$ has zero at $\g_s$ we obtain
\beq
\res_{\g_s}\Omega=<C_{n+1}^*\delta L_nC_n>\wedge \delta z(\g_s).
\label{65}
\eeq
From (\ref{h3}) and (\ref{h6}) it follows that
\beq\label{651}
<C_{n+1}^*\delta L_nC_n>
=<C_N^*\left(\prod_{m=n+1}^{N-1}L_m\right)
\delta L_n\left(\prod_{m=0}^{n-1}L_m\right)C_0>=\left(C_N^*\delta T C_0\right),
\eeq
where $T$ is the monodromy matrix. Using the standard formula for the
variation of the eigenvalue of an operator: $\delta w=C_0^*\delta T C_0$,
we obtain that
\beq
\res_{\g_s}\Omega=\delta \ln w(\g_s)\wedge \delta z(\g_s).
\label{652}
\eeq
The second set of poles of $\Omega$ is a set of branching points
$q_i$ of the cover. The pole of $C_n^*$  at $q_i$ cancels with the zero
of the differential $dz, \ dz(q_i)=0$, considered as differential on $\G$.
The vector-function $C_n$ is holomorphic at $q_i$.
If we take an expansion of $C_n$ in
the local coordinate $(z-z(q_i))^{1/2}$ (in general position when the
branching point is simple), and consider its variation, we get that
\beq
\delta C_n=-{dC_n\over dz}\delta z(q_i)+O(1).\label{66}
\eeq
Therefore, $\delta C_n$ has a simple pole at $q_i$. In the similar way,
we obtain
\beq
\delta w=-{dw\over dz} \delta z(q_i). \label{67}
\eeq
Equalities (\ref{66}) and (\ref{67}) imply that
\beq
\res_{q_i}\Omega=\res_{q_i}\left[ <C_{n+1}^*\delta L_n dC_n>
\wedge {\delta w dz\over dw}\right]\ . \label{68}
\eeq
At $q_i$ we have $dL_n(q_i)=0$. Therefore, in the way similar to (\ref{651}),
we get
\beq
\res_{q_i}\Omega=\res_{q_i}\left[ \left(C_{N}^*\delta T dC_0\right)
\wedge {\delta w dz\over dw}\right]\ . \label{681}
\eeq
Due to skew-symmetry of the wedge product we may replace $\delta T$ in
(\ref{681}) by $(\delta T-\delta w)$. Then using identities
$C_N^*(\delta T-\delta w)= \delta C_N^* (w-T)$  and
$(w-T)dC_0=(dT-dw)C_0$ we obtain
\beq
\res_{q_i}\Omega=-\res_{q_i}\left(\delta C_N^*C_0\right)\wedge \delta w dz=
\res_{q_i}\left(C_N^*\delta C_0\right)\wedge \delta w dz. \label{000}
\eeq
Note, that the $dT$ does not contribute to the residue, because $dT(q_i)=0$.

Expansions (\ref{3.9}, \ref{3.10}) near the punctures  imply
that
\beq \res_{P_{\pm}}\left(C_N^*\delta C_0\right)\wedge \delta w dz=0 . \label{001}
\eeq
Therefore,
\beq\label{h10}
\sum_{q_i}\res_{q_i}\left(C_N^*\delta C_0\right)\wedge \delta w dz=
-\sum_{s=1}^{N+1}\res_{\g_s}\left(C_N^*\delta C_0\right)\wedge \delta w dz
=\sum_{s=1}^{N+1}\delta \ln w(\g_s)\wedge \delta z(\g_s).
\eeq
The sum of (\ref{652}) and (\ref{h10}) gives (\ref{61}), because
\beq
2\omega=-\sum_{s=1}^{N+1}\res_{\g_s}\Omega-\sum_{q_i}\res_{q_i}\Omega.
\eeq
Our next goal is to prove the following statement.
\begin{lem} The symplectic form given by (\ref{54}) coincides with the canonical
symplectic structure
\beq\label{h11}
\omega=\sum_{n=0}^{N+1}\delta p_n\wedge \delta x_n.
\eeq
\end{lem}
{\it Proof.} Using the gauge transformation (\ref{3.5})
$$
L_n=g_{n+1}\tilde L_ng_n^{-1}, \ \ \C_n=g_n\tilde \C,\ \
g_n=\left(\begin{array}{cc} e^{x_n\z(z)} & 0\\
            0 & e^{-x_n\z(z)}\end{array}\right),
$$
we obtain
\begin{eqnarray}
\omega &=& \frac{1}{2}\res_{0} \ {\rm Tr} <
\tC_{n+1}^{-1}(z) \delta \tL_n(z)\wedge \delta \tC_n(z)+
\tC_{n+1}^{-1}\delta \tL_n\wedge\delta f_n\tC_n \nonumber \\
{}&-&\tC_{n+1}^{-1}\left(\delta f_{n+1}\wedge \delta \tL_n+
\delta f_{n+1}\wedge \tL_n\delta f_n\right)\tC_n > dz,
\label{h12}
\end{eqnarray}
where $\delta f_n=\delta g_n g_n^{-1}$.
From (\ref{X1}), using the equality
$$
\left(\z(x_{n+1}+x_n)-\z(x_{n+1}-x_n)\right)\delta s_{n}+
\delta (\chi_{n,-}^{1,1}+\chi_{n,-}^{1,2})=
s_{n}\delta \left(\z(x_{n+1}-x_n)-\z(x_{n+1}+x_n)\right),
$$
which follows from (\ref{3.17}), we obtain that the first term in
(\ref{h12})
$$
J_1=\res_{0} <{\rm Tr}
\left(\tC_{n+1}^{-1}(z) \delta \tL_n(z)\wedge \delta \tC_n(z)\right)>dz
$$
is equal to
$$
J_1=<{s_n\over 2s_{n+1}}\delta h_{n+1}\wedge
\delta \left(\z(x_{n}-x_{n+1})+\z(x_{n+1}+x_n)\right)>\ .
$$
Equation (\ref{X3}) implies
\beq\label{h14}
J_1=<{2\delta h_{n+1}\wedge \delta x_{n+1}\over h_{n+1}^2-1}-
{2\delta h_{n+1}\wedge\delta x_n\over h_{n+1}^2-1}
\left({\wp(x_n-x_{n+1})+\wp(x_n+x_{n+1})
\over \wp(x_n-x_{n+1})-\wp(x_n+x_{n+1})}\right)>.
\eeq
The second term in (\ref{h12}) equals
$$
J_2=<\res_{0} \ {\rm Tr}
\left(\tC_{n+1}^{-1}\delta \tL_n\wedge\delta f_n\tC_n\right) > dz=
<\res_{0} \ {\rm Tr}
\left(\tL_n^{-1}\delta \tL_n\wedge\delta f_n\right) > dz
$$
From definition (\ref{h1}) of $L_n$ by direct calculations we
obtain that
\beq\label{h16}
J_2=<{2\delta h_{n+1}\wedge \delta x_n\over h_{n+1}^2-1}
\left({\wp(x_n-x_{n+1})+\wp(x_n+x_{n+1})
\over \wp(x_n-x_{n+1})-\wp(x_n+x_{n+1})}\right)>.
\eeq
At last, the third term in (\ref{h12}) is equal to
\beq \label{199}
J_3=-<\res_{0} \ {\rm Tr}
\left(\tC_{n+1}^{-1}\delta f_{n+1}\wedge\delta \tL_n\tC_n\right) > dz=
<\res_{0} \ {\rm Tr}
\left((\delta \tL_n) \tL_n^{-1}\wedge\delta f_{n+1}\right) > dz\ ,
\eeq
because
\beq
J_4=<\res_{0} \ {\rm Tr}\left(\tC_{n+1}^{-1}\delta f_{n+1}\wedge
\tL_n\delta f_n\right)\tC_n > dz=0. \label{200}
\eeq
In order to prove (\ref{200}), let us note that at $z=0$
\beq\label{201}
f_n(z)=z^{-1}f_n^0+O(z^3),\ \
f_n^0=\left(\begin{array}{cc} x_n & 0\\ 0 & -x_n \end{array}\right).
\eeq
Therefore,
\begin{eqnarray}
J_4&=&<\res_{0} \ {\rm Tr}\left(\tC_{n+1}^{-1}\delta f^0_{n+1}\wedge
\tL_n\delta f^0_n\right)\tC_n >\wp(z) dz \nonumber \\
{}&=&<\res_{0} \ {\rm Tr}\left(\C_{n+1}^{-1}\delta f^0_{n+1}\wedge
L_n\delta f^0_n\right)\C_n >\wp(z) dz. \label{202}
\end{eqnarray}
The last term in (\ref{202}) is equal to the sum of residues at the punctures
$P_{\pm}$ of the differential
$$
<\C_{n+1}^{*}\delta f^0_{n+1}\wedge L_n\delta f^0_n\C_n >\wp(z) dz,
$$
which is holomorphic on $\G$ outside the punctures. Hence $J_4=0$.

From (\ref{199}) by direct calculations we obtain, that
\beq
J_3=<{2\delta h_{n+1}\wedge \delta x_{n+1}\over h_{n+1}^2-1}>. \label{h17}
\eeq
Equations (\ref{h14},\ref{h16}), and (\ref{h17}) imply (\ref{h11}).
Lemma is proved.

Now we are ready to complete the proof of the Theorem.
Equations (\ref{61}) and (\ref{con}) imply that
\beq\label{h70}
\omega=-\delta \a,\
\a=\sum_{s=1}^{N+1} \int_{P_+}^{\g_s} \delta \ln (\Lambda^{-N}w) dz.
\eeq
Indeed, we have
\beq\label{71}
\delta \a=\sum_{s=1}^{N+1} \delta \ln w(\g_s)\wedge \delta z(\g_s)-N
\delta \ln \Lambda\wedge \delta \sum_{s=1}^{N+1} z(\g_s).
\eeq
The last term in (\ref{71}) equals zero on the fibers $\phi_+=const$
of the map (\ref{h51}).

The differential $dS=\ln(\Lambda^{-N}w)dz$ is multi-valued on $\G$ but,
following the arguments of \cite{kp1}, one can show that
its derivatives with respect to $I_k,\ k=1,\ldots, N$ (which can be
considered as coordinates on a space of curves given by (\ref{fam1})), are
holomorphic differentials. $dS$ is an odd differential with respect to
the permutation of sheets of $\G$. Therefore, $I_{N+1}=-I_{N}$ and
the definition of $I_k$ implies that
\beq\label{81}
{\p\over \p I_k}dS=d\Omega_k^h, \ k=1,\ldots,N-1,\ \
{\p\over \p I_N}dS=d\Omega_N^h-d\Omega_{N+1}^h.
\eeq
Equations (\ref{h60}) and (\ref{ab}) imply that
\beq\label{82}
\a=\sum_{k=1}^{N-1}(\phi_k\delta I_k)+\phi_-\delta I_N,
\eeq
and using (\ref{h70}) we finally obtain (\ref{h63}). Theorem is proved.

\end{document}